# Abyss Aerosols

Authors: Xinghua Jiang[1†], Lucas Rotily[2], Emmanuel Villermaux[2,3†], Xiaofei Wang[1,4†*]

[1] *Department of Environmental Science and Engineering, Shanghai Key Laboratory of Atmospheric Particle Pollution and Prevention, Fudan University, Shanghai 200433, China*

[2] *Aix-Marseille Université, IRPHE, 13384 Marseille CEDEX 13, France*

[3] *Institut Universitaire de France, 75005 Paris, France*

[4] *Shanghai Institute of Pollution Control and Ecological Security, Shanghai 200092, China*

October 25th, 2023

†These authors contribute equally to this manuscript

*To whom correspondence should be addressed.

Correspondence to:

Xiaofei Wang: Email: xiaofeiwang@fudan.edu.cn Tel: +86-21-31242526

**Bubble bursting on water surfaces is believed to be a main mechanism to produce submicron drops, including sea spray aerosols, which play a critical role in forming cloud[1,2] and transferring various biological and chemical substances from water to the air[2–7]. Over the past century, drops production mechanisms from bubble bursting have been extensively studied[3,8–11]. They usually involve the centrifugal fragmentation of liquid ligaments from the bubble cap during film rupture[8,12], the flapping of the cap film[13], and the disintegration of Worthington jets after cavity collapse[14,15]. Here, we show that a dominant fraction of previously identified as "bubble bursting" submicron drops are in fact generated via a new mechanism underwater, inside the bubbles themselves before they have reached the surface. These drops are then carried within the rising bubbles towards the water surface and are released in air at bubble bursting. Evidence suggests that these drops originate from the flapping instability of the film squeezed between underwater colliding bubbles. This finding fundamentally reshapes our understanding of sea spray aerosol production and establishes a new role for underwater bubble collisions regarding the nature of transfers through water-air interfaces.**

Spray aerosols, such as sea spray aerosols, are the largest natural sources of atmospheric aerosols and play a critical role in affecting environment, human health, and the climate[2,8,11,16]. It is generally admitted that most atmospheric spray aerosols originate from water surface bubble bursting[12,17]. The corresponding aerosols production mechanisms have been extensively studied over the past century[6,8,11,19]. There are usually classified according to two major pathways: (1) film drops produced by the rupture of the bubble cap film[8,13]; (2) jet drops produced by the disintegration of Worthington jets after cavity collapse following bubble rupture[20–22]. A large fraction of these drops has a size less than 1 µm, those affecting much more cloud formation and radiative forcing than their supermicron counterpart[23]. Both film drop and jet drop pathways generate submicron spray aerosols[3,13,24]. Bursting of sub-100 µm bubbles do produce submicron jet drops[14], while it has recently been shown that the cap flapping instability of submillimeter bubbles produces submicron film drops[13]. We show here that a major fraction of these submicron drops released in the atmosphere cannot be produced by any of these known bubble bursting pathways.

We start with an intriguing experimental observation revealing an unusual relationship between the submicron drop production yield and the flow rate of the airstream feeding the bubbles. In this experiment, we introduce a particle-free airflow through a needle or a porous glass filter in water to produce bubbles of a specific size (Extended Data Fig. 1a). If all the drops would solely originate from bubble bursting at the liquid surface, then the number of submicron drops produced per bubble, the yield denoted *n(R)*, would primarily depend on the surface bubble cap curvature radius, *R* (Methods, Extended Data Fig. 1b). Altering the airflow rate should not affect *n(R)* as long as the bubble size remains constant. Indeed, at lower airflow rates, for bubbles with sizes in the range of 600 < $R$ < 3000 μm, $n(R)$ remains unsensitive to the airflow rate (Fig. 1a). However, upon reaching a certain airflow threshold, $n(R)$ surges by up to two orders magnitudes when the airflow rate is further increased. This surge, and the existence of an airflow threshold cannot be accounted for by standard bubble bursting.

Changing the airflow rate through the feeding needle influences the bubble formation process itself. How does this relate to $n(R)$ and why? When the airflow rate is below the threshold, each bubble forms individually, with no interaction with the ones preceding, and following it (Fig. 1b). Increasing the airflow rate causes the bubble emission frequency to increase. At some point, bubbles formed earlier collide with those formed subsequently (Fig. 1c). This defines the threshold. On occasion, these bubble-bubble collisions can even result in the creation of internal jet drops (Extended Data Fig. 2a). Thus, the most notable difference when the airflow rate reaches the threshold is the emergence of bubble-bubble collisions.

Based on the above observations, a natural hypothesis arises: Underwater bubble-bubble collisions produce submicron drops inside the bubbles, which rise towards the air/water interface and release, when their cap bursts, the drops in the atmosphere; these drops have up to now been mistakenly attributed to bubbles near bursting processes occurring at the surface[8,11], but are in fact formed early in the bubble's life. To test this hypothesis, we need to answer three specific questions: (1) Are these submicron drops indeed produced inside the bubbles? (2) If so, what is the drop production mechanism? (3) Is this mechanism relevant to actual ocean wave breaking conditions?

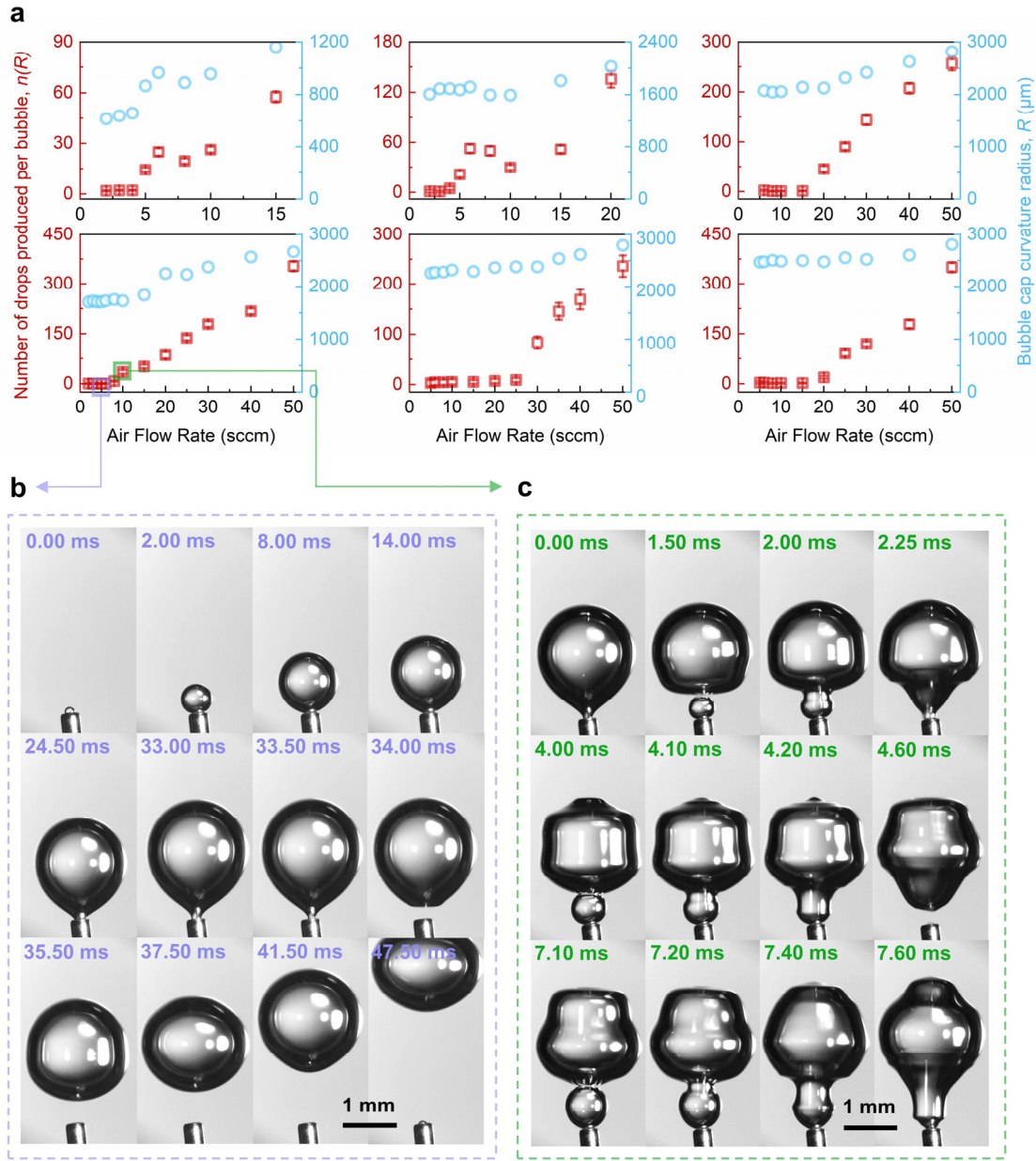

**Fig. 1: Relation between submicron drop production rate and airflow rate used to produce bubbles through a needle. a,** Number of submicron drops produced per bubble, denoted as *n(R)*, from a range of airflow rate and the corresponding bubble size *R* (Supplementary Table 1); Photos of a needle producing *R* ~ 1600 μm bubbles in water with an airflow rate at **(b)** 5 mL/min (See Supplementary Video 1) and **(c)** 10 mL/min (See Supplementary Video 2), respectively.

To investigate the first question, two experiments are designed. The first experiment relies on the fact that submicron drops inside bubbles would be captured at the ascending bubble wall via Brownian diffusion, sedimentation and other mechanisms enhanced by bubble internal circulation[25,26]. If these drops are indeed produced and suspended inside the bubbles, increasing the bubbles residence time underwater would remove part of them. For example, for air at 298 K, given that the diameter of drops inside an $R$ = 1000 μm bubble is about 100 nm, ~ 80% of drops will be removed after ~ 9 s[25]. By contrast, if the drops are produced from bubble bursting at the surface, increasing the bubble residence time underwater will not affect the drop production yield $n(R)$. We extend the residence time of bubbles by lengthening their rising distance in water (Fig. 2a) and see whether $n(R)$ changes or not. For bubbles with $R$ ~ 900, 950, 1000 and 1050 μm generated at an airflow rate of 20-50 mL/min (above the threshold for bubble-bubble collision), $n(R)$ decreases by more than one magnitude (Fig. 2b) as the underwater bubbles residence time increases, thus demonstrating that a dominant fraction of drops contributing to $n(R)$ are produced early inside the bubbles, and not after they have reached the surface.

To further confirm that underwater bubble-bubble collisions indeed produce drops, we conduct an experiment comparing the aerosols outcome in conditions with, and without colliding bubbles (Extended Data Fig. 3). Two bubble streams are produced from two needles at a distance apart, which can be varied. In one setting, the distance is approximately equal to three bubble diameters so that the bubbles form each needle are independent (the 'no collision group', Fig. 2c), while in the other setting the distance is one bubble diameter to enforce collisions between bubbles emitted by one needle and the other (termed the 'collision group', Fig. 2d). Fig. 2e shows that $n(R)$ for the collision group is at least one magnitude larger than for the no collision group, again clearly demonstrating that bubble-bubble collisions dramatically favor the production of submicron drops.

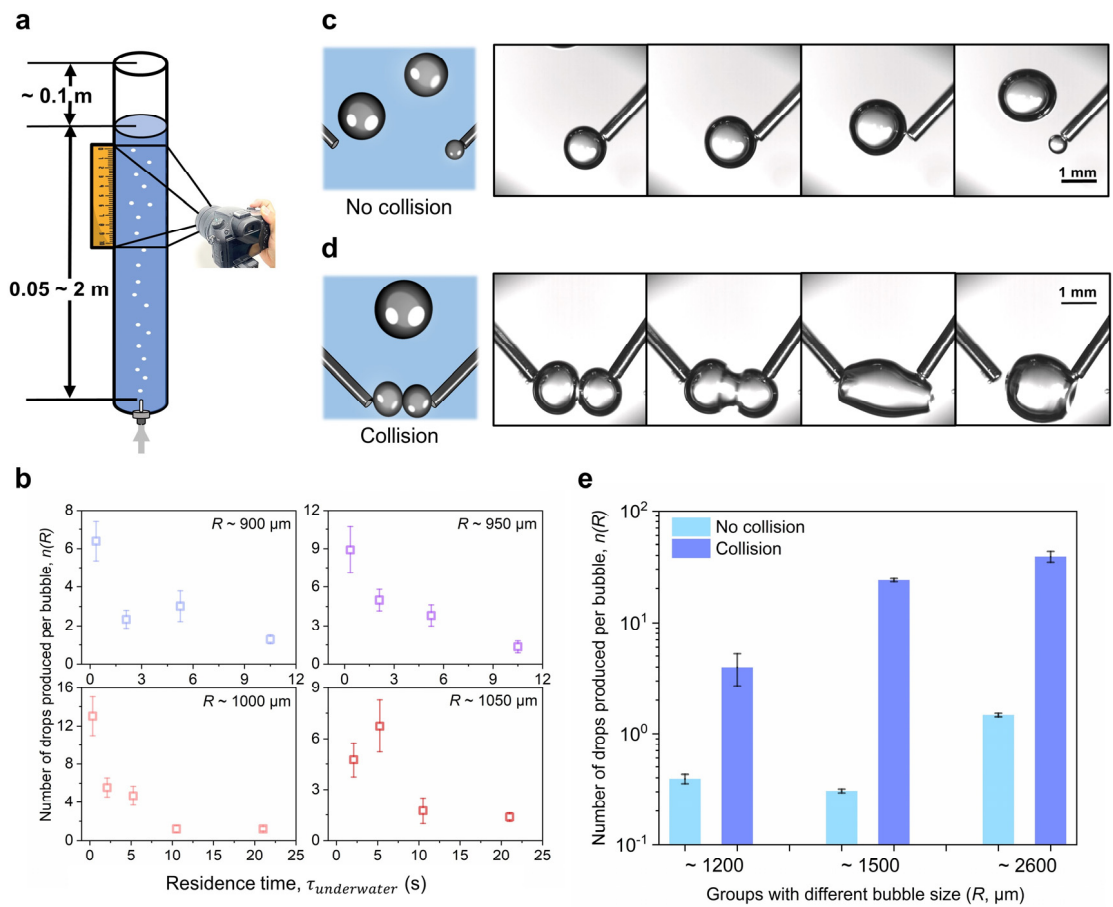

**Fig. 2: Verification of the presence of submicron drops within underwater air bubbles and the impact of bubble collisions on drop generation. a**, Schematic representation of a setup designed to prolong the underwater residence time by increasing the rising distance of bubbles in water; **b,** Number of submicron drops produced per bubble, *n(R)*, with different residence times in water; Schematic drawing and photos of bubble formation processes for **(c)** 'no collision' group (intervals of

6/1000 s, Supplementary Video 3) and **(d)** 'collision' group (intervals of 1/1000 s, Supplementary Video 4); **e**, Comparison of the number of submicron drops generated per bubble, *n(R)*, for different bubble sizes between the 'no collision' and 'collision' groups.

We now consider our second question, namely the possible mechanistic interpretations of these observations. Collisions between bubbles may produce internal jets shooting millimeter-size drops (Extended Data Fig. 2b and Supplementary Video 5). One possibility is that these jets or large drops would collide with the bubble inner wall to produce submicron splash drops. There is, however, a condition for splashing. The diameter $d$ of jet drops follows approximately the "10% rule"[3,14,18], that is $d \sim R/10$, and the velocity $v_{jet}$ of jet popping out[8] is of order

$$v_{jet} \sim \sqrt{\frac{\sigma}{\rho R}} \quad [1]$$

where $\sigma$ is the surface tension coefficient and $\rho$ is the density of water. Therefore, the Weber number $W_e$, which measures the intensity of the jet/drop impact at the bubble wall is

$$W_e = \frac{\rho v_{jet}^2 d}{\sigma} = \frac{d}{R} \sim \frac{1}{10} \quad [2]$$

way below the critical value $W_e = 200$ for splashing[27]. That option is thus ruled out.

Another possibility is that the aerosols are the fragmentation products of the film squeezed between colliding, then coalescing bubbles (Fig. 3a). When bubbles collide with a non-zero relative velocity, a liquid layer between the flattened bubbles needs to drain, and then puncture, for coalescence to occur. We simulate this process using a transparent glass tube of radius $R$, plugged at the top, partially filled with water, with air at the top as the "upper bubble" (Fig. 3b). The radius of the contact area between bubbles may vary between $0$ and a value somewhat larger than $R$ depending on the strength of the collision. We therefore use $R$ as a representative value for this radius. After slowly lifting the bottom away from the water surface, a section of air as the "lower bubble" rushes towards the water surface in the tube. After contact, the two bubbles are separated by a film which drains and finally bursts, thus completing coalescence. Surprisingly, the burst phenomenology is similar to the one known for soap films or very thin bubble caps[13,16] (Fig. 3c, Supplementary Video 6). Flapping thin films fragment in tiny drops, making of this observation a serious candidate to explain the formation of submicron aerosols from bubbles collisions. The flapping, or Squire instability onsets within a time $t_s$ such that[16]

$$t_s \sim \sqrt{\frac{\rho}{\rho_{gas}}} \frac{\sqrt{\lambda h}}{v} \qquad [3]$$

where $v = \sqrt{2\sigma/\rho h}$ is the Culick receding velocity of a film with thickness $h$ after film puncture, forming undulations of wavelength $\lambda \sim h \frac{\rho}{\rho_{gas}}$ with $\rho_{gas}$ the density of the ambient gas. For this instability to affect the film, $t_s$ should be smaller than the transit time $\tau$ of the film receding edge over the film radius $R$

$$\tau \sim \frac{R}{v} \qquad [4]$$

The condition $t_s < \tau$ implies

$$h < R \frac{\rho_{gas}}{\rho} \qquad [5]$$

The film thickness, which scales as the radius of the contact area between the bubbles,

must be thin enough for flapping to occur, which is favored in a heavier gas environment.

A film between two bubbles of comparable sizes is typically flat. But it drains for the same reason the curved cap of a bubble at the surface of a pool does: The pressure in a bubble cap with radius of curvature $R$ is[8]

$$p = p_0 + \frac{2\sigma}{R} \qquad [6]$$

where $p_0$ is the external pressure. A flat film squeezed between two bubbles has pressure $p_0$, but the pressure at the border of the contact area (the tube radius in our analogue experiment) is

$$p_0 - \frac{\sigma}{R} \qquad [7]$$

In both cases, the liquid is pressurized with respect its environment, where it empties, thins until it finally ruptures[28,29]. The thinning dynamics (see SI) leads to

$$h(t) \sim R \left(\frac{t}{t_v}\right)^{-\frac{2}{3}} \qquad [8]$$

where $t_v = \frac{\eta R}{\sigma}$ is the viscous time, with $\eta$ the liquid viscosity. The bursting time corresponds to the contamination time at the film edge[29]

$$T \sim \frac{h}{u} Sc^{\frac{2}{3}} \qquad [9]$$

where $Sc = \nu/D$ (with $\nu = \eta/\rho$ and $D$ is mass diffusivity) is the Schmidt number of the impurities feeding the marginal regeneration responsible for the film integrity, and drainage dynamics at velocity $u \sim \frac{\sigma}{\eta}\left(\frac{h}{R}\right)^{\frac{3}{2}}$. The bursting condition $T < R/u$ provides the bursting time $t_*$ as

$$t > t_v Sc = t_* \gg t_v \qquad [10]$$

and the film thickness at burst

$$h(t_*) = RSc^{-\frac{2}{3}} \tag{11}$$

Combining equation [11] with [5], we see that a ruptured film will spontaneously flap provided

$$Sc^{\frac{2}{3}} \frac{\rho_{gas}}{\rho} > 1, \tag{12}$$

a condition, with $Sc = O(10^4)$, always fulfilled in air (where $\frac{\rho_{gas}}{\rho} = O(10^{-3})$). Interestingly, this condition puts no constraint on the colliding bubbles contact area $R$, meaning that whatever the strength of the collision may be ($R$ is larger for a larger bubbles collision velocity), flapping is likely to occur identically.

The size of the drops $d$ thus produced scales like, but is typically much smaller than $R$[11,13]

$$d \sim \sqrt{vt_s h} \sim h \sqrt{\frac{\rho}{\rho_{gas}}} \sim RSc^{-\frac{2}{3}} \sqrt{\frac{\rho}{\rho_{gas}}} = O\left(\frac{R}{10^3}\right). \tag{13}$$

This rule predicts that the film drops produced by millimeter-size or sub-millimeter colliding bubbles will be submicronic, which is consistent with our experimental observations (Extended Data Fig. 4).

The above theory predicts that flapping would occur earlier, and would therefore be more efficient at producing aerosols, in a gas with a higher density[13], which is confirmed by Fig. 3d, showing that flapping during bubble collision in an SF$_6$ environment has a significantly higher magnitude (Supplementary Video 7), resulting in more film fragmentation than in air.

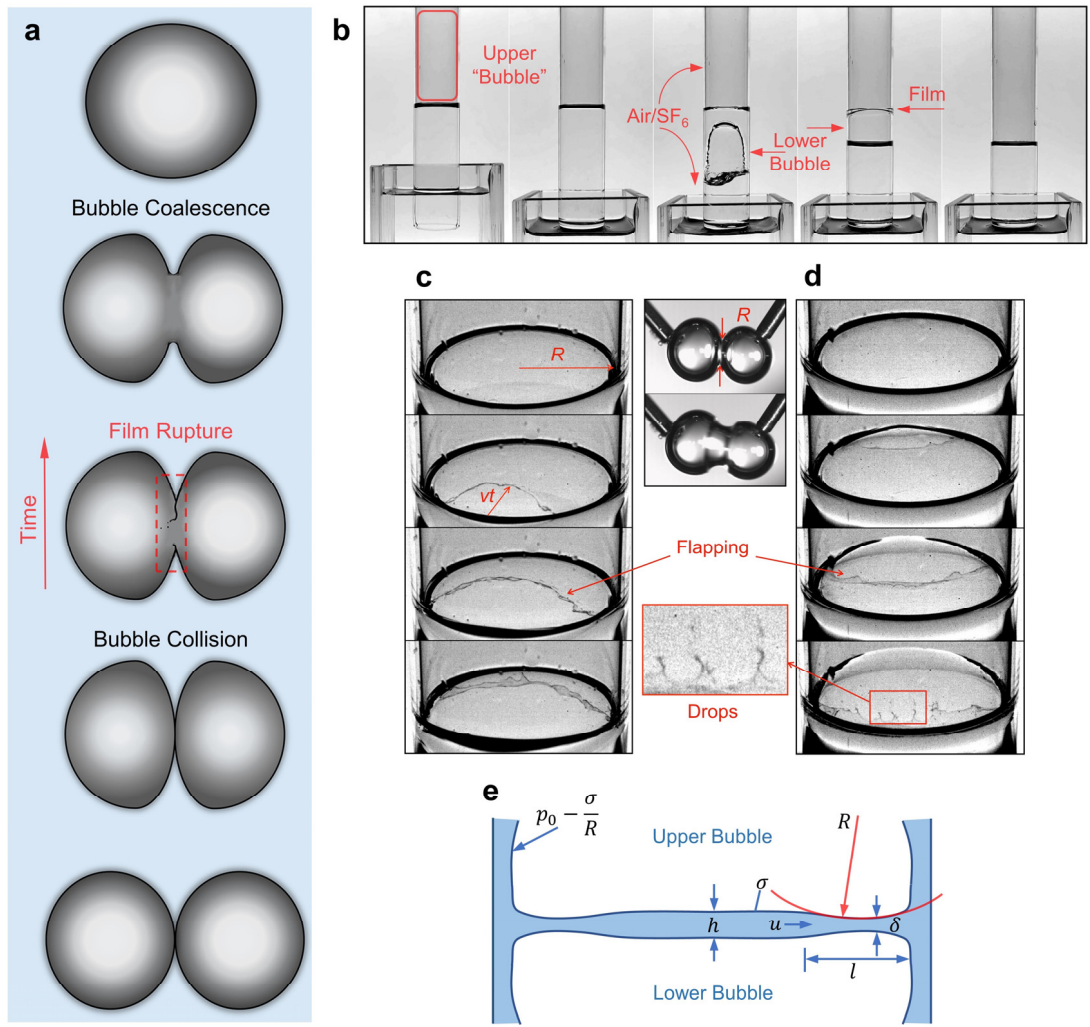

**Fig. 3: Film fragmentation and drop production during bubble collision. a**, Schematic illustration of the proposed film fragmentation process during an underwater bubble collision; **b,** Analogue of a bubble-bubble collision in a glass tube (Supplementary Video 8); Flapping film at burst during a bubble-bubble collision (**c**, air-water interface; **d**, $SF_6$-water interface), taken at intervals of 1/2250 seconds (Supplementary Video 6, 7) showing liquid ligaments, and drops. The inset of **c** shows $R$ of the contact area between bubbles, which is analogous to $R$ in glass tube; **e**, Schematic diagram of the film and parameters ruling its drainage before rupture.

Fig. 4 summarize the film drop production mechanism. Comparing to surface bubble bursting, underwater bubble collision is a much more efficient mechanism to produce submicron drops. It does not require a specific bubble size range. However, submicron drops in sub-100 μm bubbles would be removed by diffusion very quickly. For example, ~51% of 160 nm drops will be removed in a $R$=100 μm bubble within just 0.1 seconds (SI). Thus, these small bubbles would not be able to release submicron drops from underwater collisions, which is consistent with the data shown in Extended Data Fig. 5.

Sea spray aerosol, the most important natural spray aerosol in the atmosphere, are created by bubbles from wave breaking. Thus, our third question is whether the new drop production mechanism plays a major role in forming sea spray aerosols. To obtain a rough estimate, we used a plunging jet, a common analog of wave breaking in the laboratory[30–32], to simulate bubble and sea spray aerosol production. There are obviously many bubbles collisions in a plunging jet (Supplementary Video 9). The collisions induce bubbles deformation amplitudes similar to gravity-induced deformations of surface bubbles for collision relative velocities of the order of $\sqrt{gR}$, that is 0.1 m/s for a millimetric bubble, a value largely met in ocean wave breaking[33].

We point a high-speed camera to an 8.0×8.0×0.25 mm³ water volume in a bubble forming region (close to the incipient jet) and find that ~8.1% bubbles with $R$>~200 μm have experienced collisions with each other in this volume during a 0.35 s period (Supplementary Video 9). These bubbles may also have collisions outside this region and time. Thus, the actual percentage of bubbles which have experienced collisions should be much larger than 8%. Fig. 2e shows that bubble collisions are at least 10 times more efficient than bubble bursting. Moreover, bubbles with $R<\sim100$-$200$ μm are likely to be trapped by the turbulent fluctuations in ocean and cannot reach sea surface[34]. Thus, for wave breaking bubbles with $R>\sim200$ μm, the number of submicron drops produced from underwater bubble collision is likely to be greater than those from bubble bursting.

The effect of collisions is of course also obvious on the yield *n(R)* in wave breaking bubbles. The submillimeter bubbles generated by a plunging jet are sucked into another container and burst (Extended Data Fig. 6). It is found that the bubbles with a size range of $200 < R < \sim700$ μm produce ~2 drops per bubble, at least 5 times more than those produced by bubble bursting only (Fig. 4), demonstrating that bubble collisions in this size range is a major pathway to produce submicron drops.

Natural bodies of water, such as oceans and lakes, cover most of the Earth's surface[2,35]. Drops produced from bursting bubbles, usually generated by wave breaking, are a major source of atmospheric aerosols, as well as a primary medium for material transfer at the water-air interface[10,36]. We have demonstrated that aerosol production is not solely confined to the surface of water, but that it should be seek for in the abyss: Underwater bubble collisions are the cause of most of submicron drops, which were before mistakenly regarded as bubble bursting aerosols. The production flux of sea spray aerosol is thus dependent on not only bubbles sizes distribution, but also on the bubble collision frequency, a parameter never being considered in current global climate models. Material transfer across the water-air interface does not only occur at water surfaces, but also at underwater bubble walls. This finding fundamentally reshapes our understanding of spray aerosol production and invites to reconsider the mechanisms form which they originate, notably bubbles collisions, in all the relevant environmental and industrial processes where they arise.

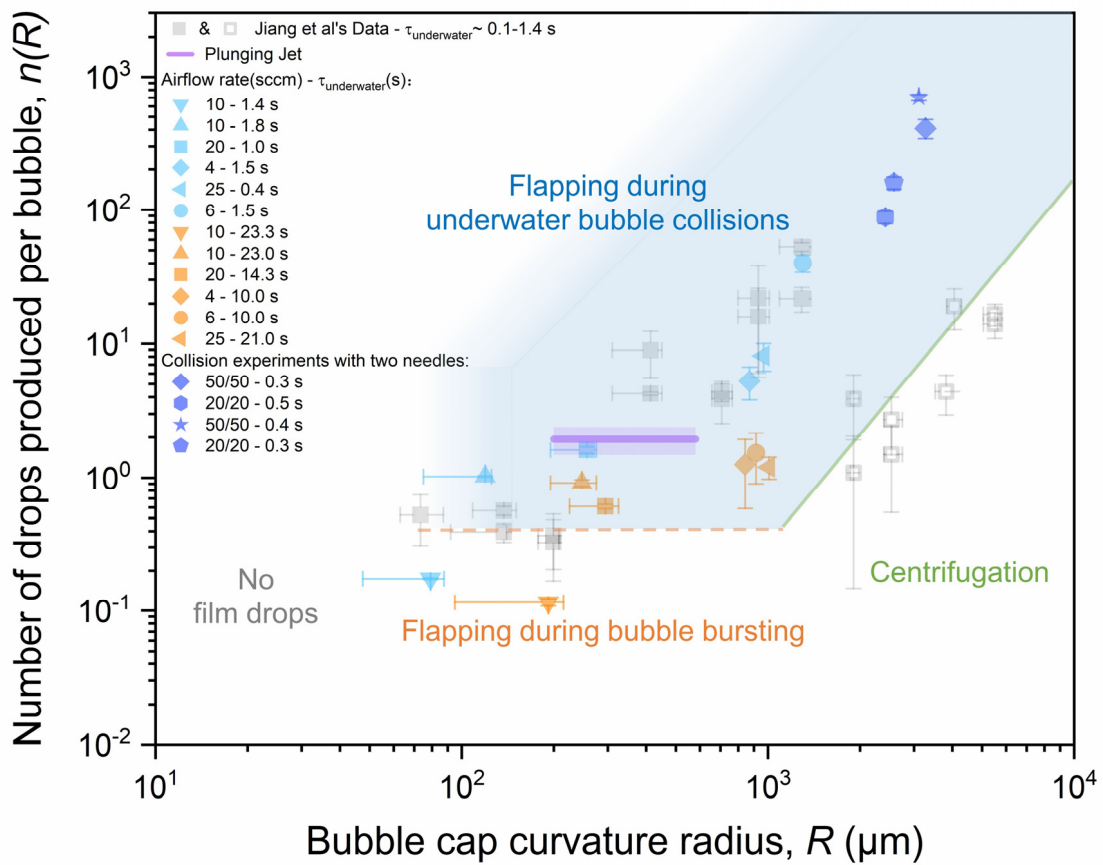

**Fig. 4: Production mechanisms of submicron drops.** The dashed orange line and solid green line depict the yield $n(R)$ resulting from the flapping mechanism and centrifugation mechanism, respectively, during surface bubble bursting. The light blue region represents the possible $n(R)$ range for underwater bubble collision. The purple horizontal line represents $n(R)$ generated by plunging jet bubbles with $200 < R < \sim 700$ μm (Extended Data Fig. 7); The gray square points represent data reported in Jiang et al.[13] The solid markers indicate the presence of bubble collisions, as the airflow during bubble generation surpasses the established threshold, while the hollow markers signify the absence of collisions. $\tau_{underwater}$ is the residence time of bubbles in water, for different

airflow rates through the needle or porous glass filter to produce bubbles. The error bars in bubble size were derived from the abscissa of the cumulative frequency of ∼25 to 75% in the frequency distribution diagram (Extended Data Fig. 8).

# Methods

**Experimental water**

The water utilized in this study is a 3.5% sodium chloride solution. This solution is prepared using ultrapure water with a resistivity of 18 MΩ·cm and NaCl (analytically pure, China National Pharmaceutical Group Co., Ltd.).

**High-speed camera imaging**

For the acquisition of images depicted in Fig. 1 **b**, **c** and Fig. 2 **c**, **d** as well as Supplementary Video 1-5, 9, a high-speed camera (Fastcam SA-Z 1000K-M-32GB, Photron. Ltd) equipped with a 2.5-5x magnification macro lens (FF 25mm F2.8 Ultra Macro 2.5-5x, Anhui Changgeng Optics Technology Co., Ltd) is employed. The camera is set to capture at a rate of 20,000 frames per second (fps), with each frame having an exposure time of 1/20664 s. The photos in Fig. 3 **c, d** are snapshots from videos taken with a Photron Fastcam Mini AX200 type 900K-M-32GB fitted with a Nikon macro lens.

**Experimental setup**

*A. Experimental setup for measurement of aerosol concentration and size distribution*

The setup is shown in Extended Data Fig. 1a. A zero-air generator produces a particle-free air flow, which is split into two streams, regulated by mass flow controllers (MFC, Beijing Sevenstar Flow Co., Ltd.). One stream is directed through a glass filter or needle to produce bubbles of varying sizes. Concurrently, the other stream is introduced directly into a tank headspace at a specific flow rate (Supplementary Table 1). Prior to entering any aerosol instruments, silica diffusion dryers are always utilized to dry these drops into solid particles. To prevent any potential ambient aerosol contamination within the system, a rigorous purging process is undertaken using particle-free air. This purging continues until the particle concentration reaches a negligible threshold (< 0.01 particles/cm$^3$). Particle concentrations are quantified using a condensational particle counter (CPC) (Model 3775, TSI Inc.). An impactor (with cut-off particle size of ~1 μm at 0.3 LPM) is placed before the CPC, removing most supermicron particles.

Moreover, the particle size distributions are measured by a scanning mobility particle sizer (SMPS) system, which comprises of two main components: a differential mobility analyzer (Model 3081, TSI Inc.) and a CPC (Model 3775, TSI Inc.). Some typical size distributions of dried drops produced from bubbles in this setup are shown in Extended Data Fig. 4.

B. *Setup modification for extending underwater bubble residence time*

The setup modification is shown in Fig. 2a. The system is designed to produce bubbles via a glass filter at the base of a tube. The cylindrical tube has an internal diameter of 5.5 cm. Its length is sufficient to fill different liquid levels required for the experiment, as detailed in Supplementary Table 1. Notably, there is an approximate 10 cm distance from the water surface to the device's lid. The rising velocities of bubbles are measured by the slow-motion movie captured by a camera at 240 fps. The residence time of bubbles in water can be calculated. The particle size distributions for bubbles with $R =$ ~900 μm from this experimental setup are shown in Extended Data Fig. 9.

*C．Setup for visualizing flapping film during the simulated bubble collision.*

The setup is depicted in Fig. 3b (refer to Supplementary Video 8). A glass tube with an internal diameter of roughly 2 cm is submerged in a 3.5% sodium chloride solution. The solution is drawn up to a height of about 5 cm above the liquid level, after which the tip of the tube is sealed. Carefully, the tube is raised until its base was level with the surface of the liquid. A gentle tilting motion is given to the tube, which is then quickly realigned, allowing ambient air to enter and create a rising bubble. As this bubble have ascended, it collides with the bubble above, resulting in a film's formation. A high-speed camera captures the complex dynamics of the film's flapping and rupture, shedding light on the patterns of disintegration during the breaking phase.

*D．Setup modification for studying bubble collision*

The setup is depicted in Extended Data Fig. 3. It consists of a transparent square-shaped acrylic box, with dimensions of 20×5×20 cm. The carrier gas stream maintains a steady flow rate of 1 LPM to the headspace, and two other streams, both with equal volume flow rates, remain constant (refer to Supplementary Table 1). Air is channeled through two identical needles to produce bubbles, with the outlets positioned 5 cm beneath the water's surface. Consequently, the resulting residence time is approximately 0.2 seconds, ensuring the effects of drop removal within the bubbles are minimal. To study the impact of underwater bubble collisions on drop production, two separate control experiments are designed based on the spacing of the outlets: In the 'Collision' group, the two outlets are spaced by one diameter of the generated bubble to ensure two bubbles colliding with each other; for the 'No collision' group, the spacing is approximately equal to three times of the bubble's diameter. To help prevent potential collisions between consecutive bubbles and the ones preceding them, the needles are angled downward at 45 degrees.

*E．Experimental setup for quantifying n(R) for plunging jet bubbles*

The setup is shown in Extended Data Fig. 6. The experimental design incorporates two identical cylindrical acrylic tanks, each with an internal diameter of 24 cm and a height of 30 cm. Both tanks are filled with 8 L of a 3.5% sodium chloride solution. Within Tank A, the primary peristaltic pump (Model: F6-12L YZ35, Innofluid Co., Ltd.) is employed to extract the salt water at a constant flow rate ($Q_{jet}$), subsequently reintroducing it via a 1/4-inch Teflon tube positioned at the tank's lid to produce a jet towards water surface, thereby generating bubbles like wave breaking.

Another peristaltic pump (Model: BT-300CA 253Yx, Jieheng Peristaltic Pumps Co.,

Ltd.) is utilized to draw the saline solution from Tank A, which contains bubbles produced by the plunging jet, and then directly transfer it to Tank B at a flow rate of 0.6 LPM ($Q_{suction}$). A connecting conduit is integrated between the two tanks, ensuring a stable water level for both tanks. The particle concentrations in Tank B headspace are measured by a CPC.

Furthermore, to measure the size distribution of bubbles drawn into and subsequently ruptured within Tank B, a segment of a square, flat, transparent pipe is incorporated into the suction pipeline, mitigating distortions from convex lens refraction inherent in circular conduits. A mechanical shutter camera (Canon EOS R6 Mark II with a 5x magnification lens) is deployed, boasting an exposure time of 1/8000 s, to capture the dynamics of the bubbles in transit. This facilitates the measurement of bubble diameters and the subsequent calculation of the actual frequency size distribution of the sucked bubble (Extended Data Fig. 7).

**Quantification of drop production from $R > 200$ μm bubbles in plunging jet.**

To quantify drop production from bubbles with $R > 200$ μm within plunging jets, a significant challenge arises as bubbles with $R > 200$ μm mix with sub-200μm bubbles. To address this challenge and minimize interference from these sub-200μm bubbles, we devise two suction settings tailored to different bubble size ranges: The Downward suction positioned proximal to the jet and primarily targets the extraction of all bubbles

(larger bubbles with $R > 200$ μm and sub-200μm bubbles); while the Upward suction is designed to only transport sub-200μm bubbles located at the tank's periphery, serving as a background control. During the suction process, the bubble size distribution is measured by vertically photographing the flat channel, and $n(R)$ for bubble with $R > 200$ μm is obtained. The Downward suction captures more bubbles with $R > 200$ μm compared to the Upward group. The difference is:

$$N_{D,R>200\ \mu m} = N_{Downward} \times f_{Downward} - N_{Upward} \times f_{Upward} \times \frac{N_{Downward}}{N_{Upward}}$$

where, $N$ is the total number of bubbles suctioned per time unit; $f$ is the fraction of bubbles with $R > 200$ μm relative to the total bubble population. The subscripts 'Downward' and 'Upward' denote the experimental groups subjected to Downward and Upward suction, respectively. All the values on the right side of the equation are obtained by analyzing the actual frequency size distribution of the sucked bubble (Extended Data Fig. 7).

The number of drops produced by these bubbles ($N_{D,R>200\ \mu m}$) is:

$$n_{D,R>200\mu m} = Q\ c_{Downward} - Q\ c_{Upward}\frac{N_{Downward}}{N_{Upward}}$$

where, $c$ is the dried drop concentration measured by the CPC; $Q$ is the flow rate of the particle-free carrier air. Thus,

$$Drops\ per\ bubble: n(R > 200\ \mu m) = \frac{n_{D,R>200\mu m}}{N_{D,R>200\ \mu m}}$$

**Measurement of bubble cap curvature radius $R$.**

To measure the size of bubbles, two distinct methodologies were employed based on the radius $R$. For bubbles with $R < \sim 200$ μm, bubbles were imaged underwater using a high-resolution microscope camera (TipScope CAM model, TipScope Inc.). Since bubbles with $R < \sim 200$ μm maintain a spherical shape when submerged, we can deduce the equivalent volume radius $R_{equ}$ by comparing the diameter of the in-focus bubble to a known scale.

For bubbles with $R > \sim 200$ μm, we use slow-motion videography to count the bubbles produced within a specific time frame. By dividing the volume of gas used to generate these bubbles by their count, we determine the average volume of a single bubble. We then calculated the equivalent volume radius $R_{equ}$ using the formula for the volume of a sphere: $V = \frac{4}{3}\pi R_{equ}^3$.

The volume equivalent radius can be converted to the bubble cap curvature radius $R$, based on Toba (1959) [37].

**Measurement of the total number of bubbles with $R < \sim 200$ μm rising a long distance.**

The conventional method for determining the total bubble count involves dividing the airflow rate used to produce the bubbles by the average volume of a single bubble. However, for tiny bubbles with $R < \sim 200$ μm, a significant portion of the bubble's air dissolves in the water before it reaches the surface. As a result, the volume of bubbles that actually reach the surface is different from the volume of air used to create them. This discrepancy makes the flow rate-based calculation of the bubble count inaccurate.

To rectify this, we have devised an experimental setup to more accurately measure the volume of bubbles that burst upon reaching the water surface (see Extended Data Fig.10). As bubbles ascend, some are captured by an inverted test tube positioned just above the water. When these bubbles burst, the liquid level inside the tube is lowered. We carefully recorded the time it took for a 10 mL decrease in this liquid level. By comparing the area of the test tube's opening to the entire water surface area, we can infer the volume flow rate of the bubbles that burst at the surface. This technique offers a much more precise estimate of the number of bubbles that rise to the water's surface.

**Definition of $D_p$.**

In this paper, before conducting aerosol measurements, all drops are passed through silica diffusion dryers to transform them into dried particles. Given that the water's salinity is 3.5%, the diameter $D_p$ of the dried particle is approximately 1/4 of the original droplet diameter. The particle diameter $D_p$ is often determined by its measurement method. Both Extended Data Fig. 4, 5b, 9 were obtained using Scanning Mobility Particle Sizer (SMPS) measurements. As the SMPS measures the electrical mobility particle diameter ($D_m$), we define the particle diameter $D_p$ in this article as

$D_m$.

## Data availability

All the experimental data and videos in this work will be available online once the manuscript is accepted by a peer-reviewed journal.

## Acknowledgement

This work was supported by the National Natural Science Foundation of China (Nos. 42377090 and 42077193), the Shanghai Natural Science Foundation (23ZR1479700), and the National Key R&D Program (No. 2022YFC3702600 and 2022YFC3702601).

## Contributions

XJ, EV and XW contributed equally to this work. XW identified the scientific question and initiated this study. EV conducted the theoretical analysis and performed simulated bubble collision experiments in test tube with the help of LR. XJ performed the aerosol and bubble size distribution measurement, recorded bubble collision with a high-speed camera and conducted data analysis. All authors contributed to manuscript writing.

## Corresponding author

Correspondence to Xiaofei Wang.

## Ethics declarations

**Competing interests:** The authors declare no competing interests.

# Supplementary Information for

# Abyss Aerosols

Authors: Xinghua Jiang[1†], Lucas Rotily[2], Emmanuel Villermaux[2,3†], Xiaofei Wang[1,4†*]

[1] *Department of Environmental Science and Engineering, Shanghai Key Laboratory of Atmospheric Particle Pollution and Prevention, Fudan University, Shanghai 200433, China*
[2] *Aix-Marseille Universite, IRPHE, 13384 Marseille CEDEX 13, France*
[3] *Institut Universitaire de France, 75005 Paris, France*
[4] *Shanghai Institute of Pollution Control and Ecological Security, Shanghai 200092, China*

†These authors contribute equally to this manuscript

*To whom correspondence should be addressed.

Correspondence to:

Xiaofei Wang: Email: xiaofeiwang@fudan.edu.cn Tel: +86-21-31242526

**This PDF file includes:**

Supplementary Information Text
Extended Data Figure 1-10
Supplementary Table 1-2
Captions for Supplementary Video 1-9

# SI TEXT

## Theoretical derivation of bubble film thickness ($h$) versus time

The equations describing the quasi-steady drainage of thin liquid film emptying at the wetted wall of a tube with radius $R$ are[1–3]:

$$\frac{h-\delta}{l^2} \sim \frac{1}{R} \quad (Curvature/pressure\ adaptation) \qquad [S1]$$

$$\frac{\eta u}{\delta^2} \sim \frac{\Delta p}{l} = \frac{\sigma}{Rl} \quad (Viscous\ pressure\ loss - Stokes\ equation) \qquad [S2]$$

$$\frac{\eta u}{\delta} \sim \frac{\Delta \sigma}{l} \quad (Marangoui\ stress) \qquad [S3]$$

where $\eta$ is liquid viscosity, and $\Delta p$ and $\Delta \sigma$ are the pressure difference and surface tension difference between the film and the meniscus. Typically, $\delta \lesssim h$, therefore $l \sim \sqrt{Rh}$ so that, with $\frac{\eta u}{\delta} \sim \frac{\sigma}{R}\frac{\delta}{l}$, we have

$$\Delta \sigma = \sigma \frac{\delta}{R} \qquad [S4]$$

The equations above are quasi-steady (no time-derivatives) meaning that the film thins self-similarly:

$$\frac{(h\dot{R}^2)}{hR^2} = \frac{(Rl\dot{\delta})}{Rl\delta}, \qquad [S5]$$

Meaning that all sub-volumes empty at the same rate. Thus,

$$\frac{\dot{h}}{h} \sim \frac{\dot{\delta}}{\delta} \qquad [S6]$$

And we get $\delta \lesssim h$. Thus,

$$\frac{\eta u}{h^2} \sim \frac{\sigma}{R\sqrt{Rh}}, \qquad [S7]$$

Which leads to the interstitial emptying velocity

$$u \sim \frac{\sigma}{\eta}\left(\frac{h}{R}\right)^{\frac{3}{2}}. \qquad [S8]$$

Finally, from the global mass balance:

$$\pi R^2 \dot{h} = -u 2\pi R h, \qquad [S9]$$

we obtain the dynamics of the film thickness:

$$\frac{\dot{h}}{h} = -\frac{2u}{R} = -\frac{2}{R}\frac{\sigma}{\eta}\left(\frac{h}{R}\right)^{\frac{3}{2}}, \quad [S10]$$

leading to its evolution as a function of time:

$$h(t) \sim R \left(\frac{\sigma}{\eta}\frac{t}{R}\right)^{-\frac{2}{3}} \quad [S11]$$

**Calculation of drop removal rate in a sub-100 μm bubble**

The calculation only accounts for drop loss by diffusion on the bubble inner wall. Thus, the actual drop removal rate should be larger than the calculated value, as other mechanisms, such as gravitational settlement. The equation used for the calculation here is[4]

$$\frac{N(t)}{N(0)} = \frac{3}{8}\sum_{p=1}^{\infty} A_n^2 \exp\left(-\mu_p \frac{16Dt}{R_{equ}^2}\right) \quad [S12]$$

Where $D$ is the diffusion coefficient; $A_n$ and $\mu_p$ are constants: $\mu_1 = 1.678$, $\mu_2 = 9.83$, and $A_1 = 1.32$, $A_2 = 0.73$. Notably, this calculation can only give a lower bound of the drop removal rate.

# EXTENDED DATA FIGURES

## Extended Data Fig. 1

**a**, Schematic drawing of the experimental setup used to measure spray aerosol from bubbles at various airflow rates. The glass filter can be replaced by a needle. **b**, Schematic drawing of the bubble cap curvature radius $R$ and bubble equivalent radius $R_{equ}$. Both can be converted to each other[5,6].

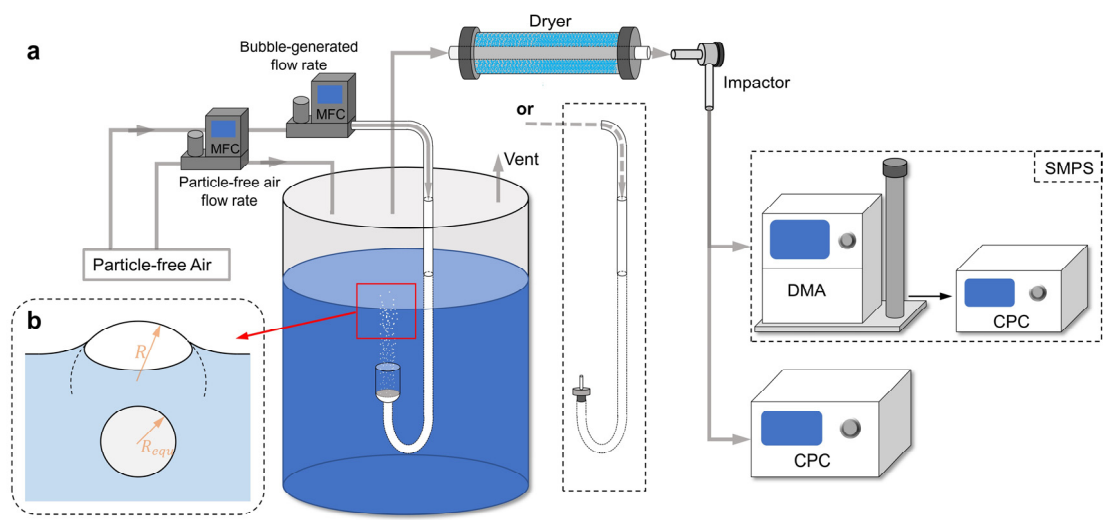

**Extended Data Fig. 2**

**a**, Photo of internal jet drops produced by bubble-bubble collisions (from Supplementary Video 2); **b**, Photo of internal jets and millimeter size drops produced by bubble-bubble collision (from Supplementary Video 5).

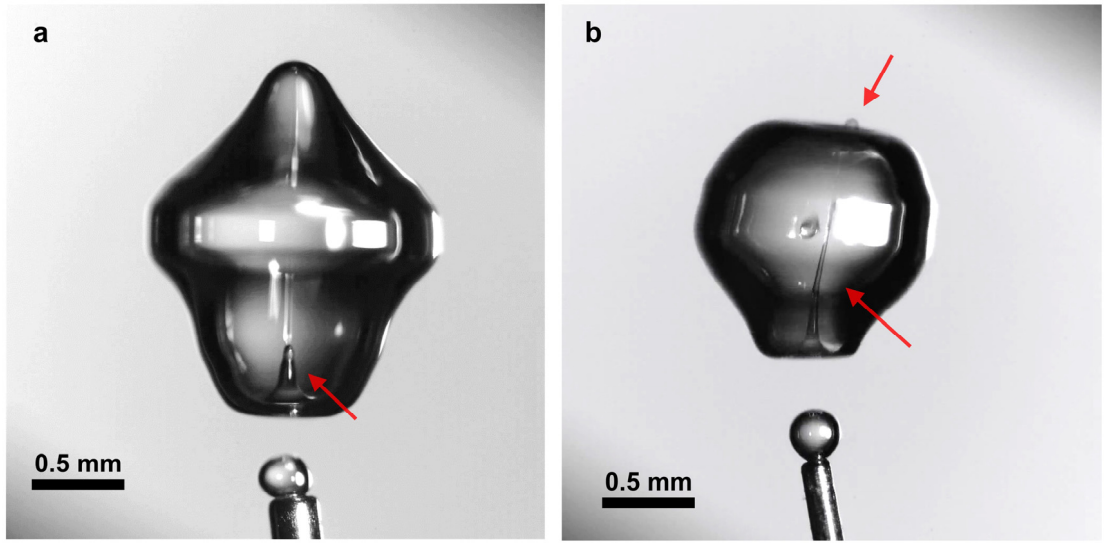

**Extended Data Fig. 3**

Schematic drawing of the setup for the underwater collision experiment. The distance of two streams of bubbles decides whether bubbles collide with each other or not.

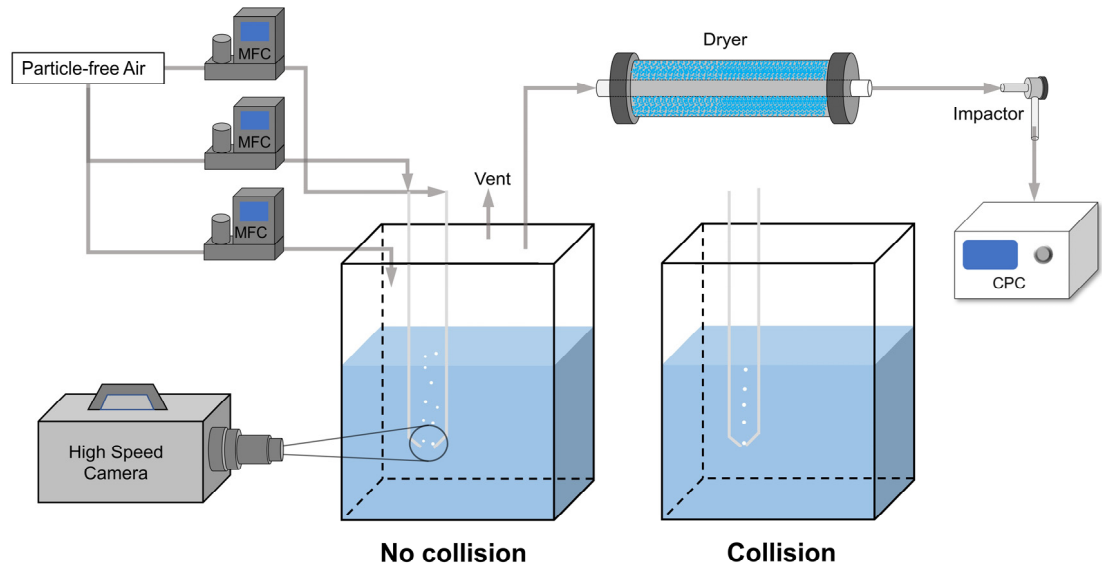

**Extended Data Fig. 4**

Particle size distributions of dried drops produced from the bubbles with different sizes in terms of probability (P) density function. The airflow rates exceed the threshold for bubble collision, and the residence time $\tau_{underwater}$ of all these bubbles are less than 1 second.

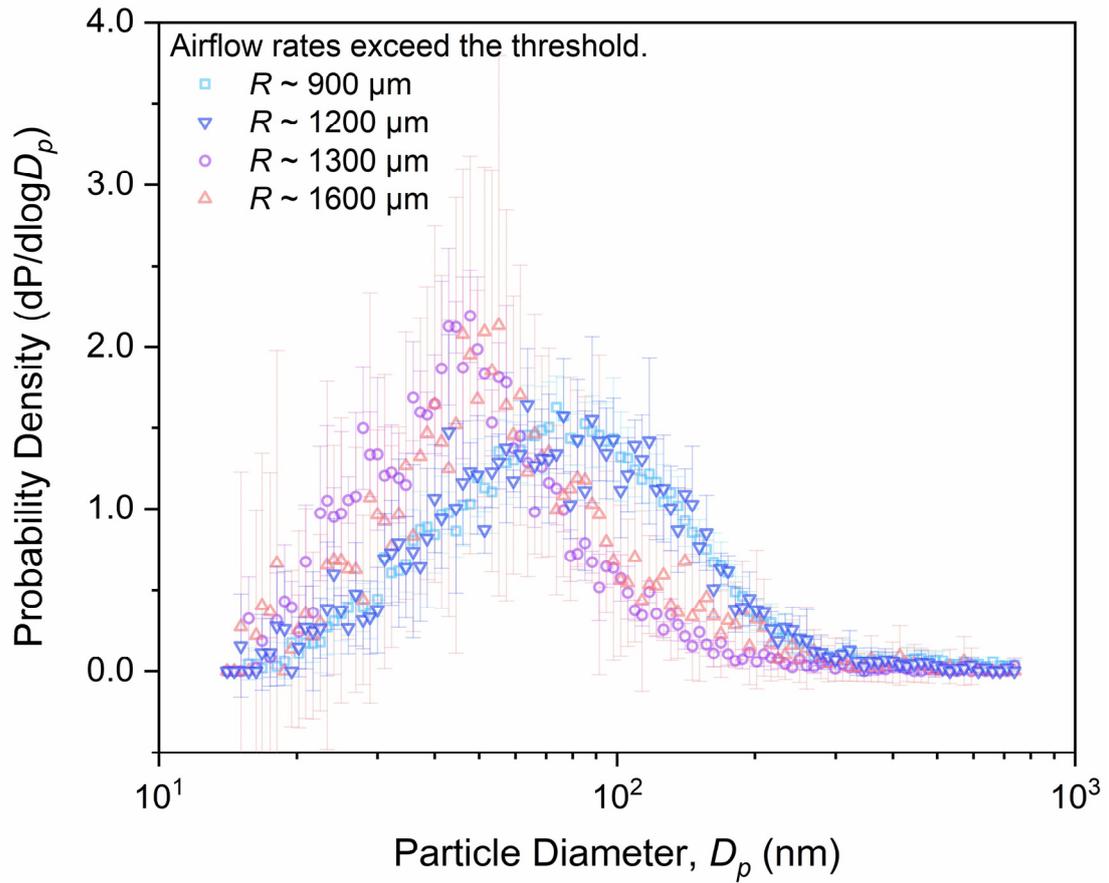

**Extended Data Fig. 5**

**a**, Number of submicron drops produced per bubble $n(R)$ ($R \sim 80$ μm) with a range of air flow rate at different underwater residence time; **b**, Particle size distribution produced by $R \sim 80$ μm bubbles with residence time of ~5.6 s at different air flow rates. Obviously, the $n(80\ \mu m)$ does not change significantly with both air flow rates and underwater residence time.

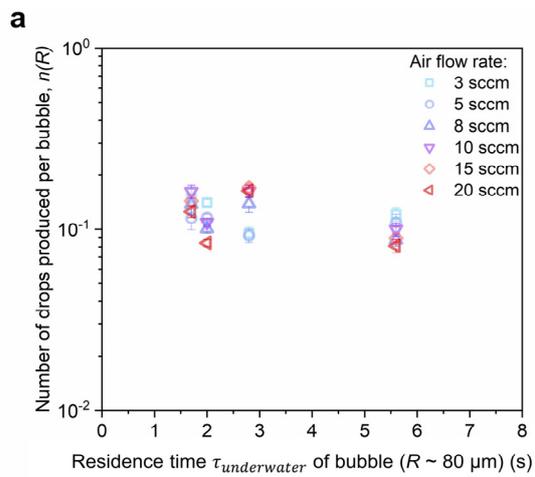
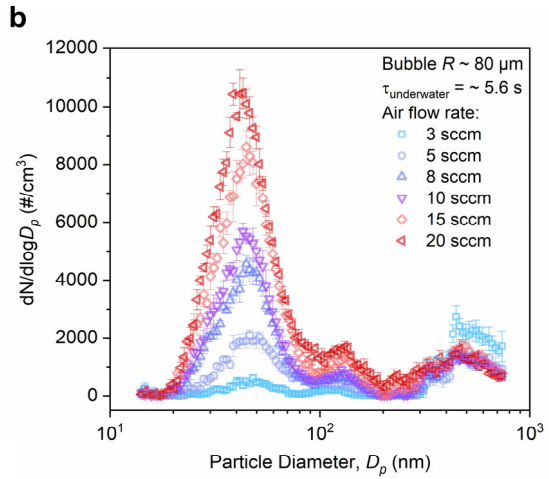

**Extended Data Fig. 6**

Schematic diagram of the experimental setup for extracting plunging jet bubbles of different size ranges into another container for bursting by two different methods: (1) Downward vertical suction, which can capture more bubbles with $R > 200$ μm; (2) Upward vertical suction, which can capture fewer bubbles with $R > 200$ μm.

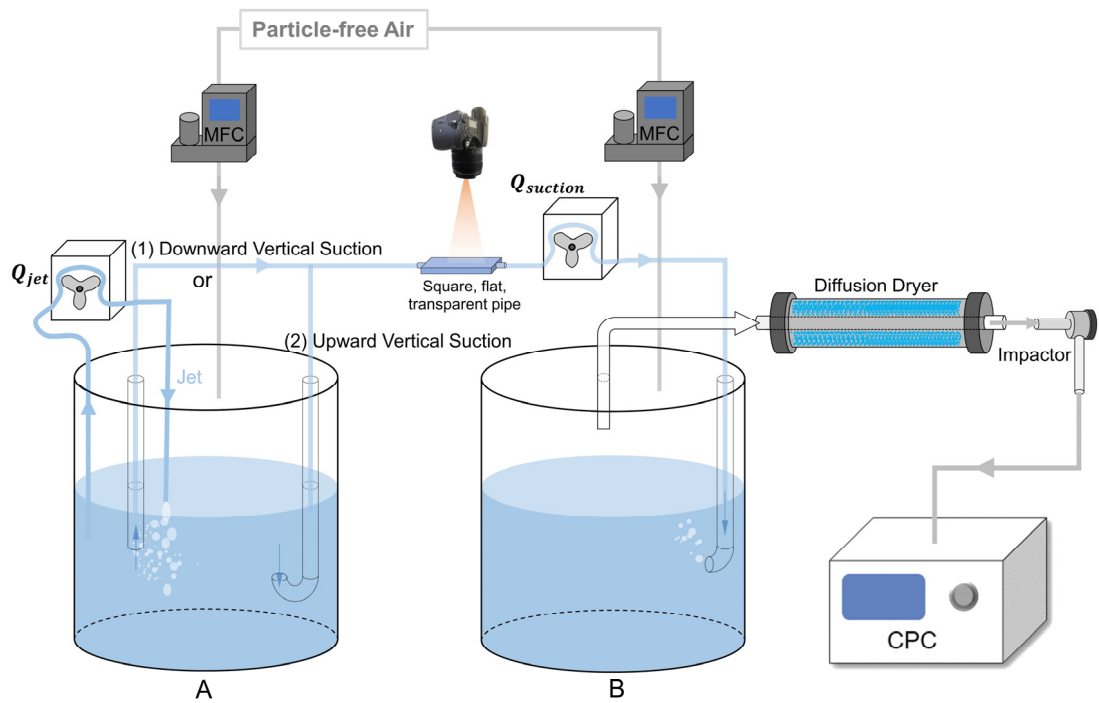

**Extended Data Fig. 7**

Bubble size distributions of the extracted plunging jet bubbles. The plunging jet impacts the water surface at a flow rate of 2.25 LPM. The extractions are conducted through vertically downward or upward suctions.

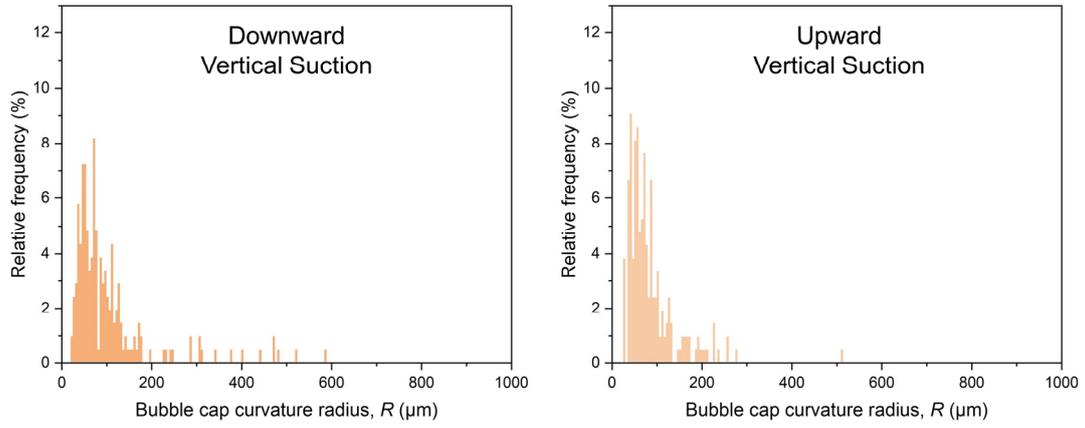

**Extended Data Fig. 8**

Size distributions of bubbles generated by the air passing through glass frits with different porosities. See Supplementary Table 2 for the detailed experimental conditions.

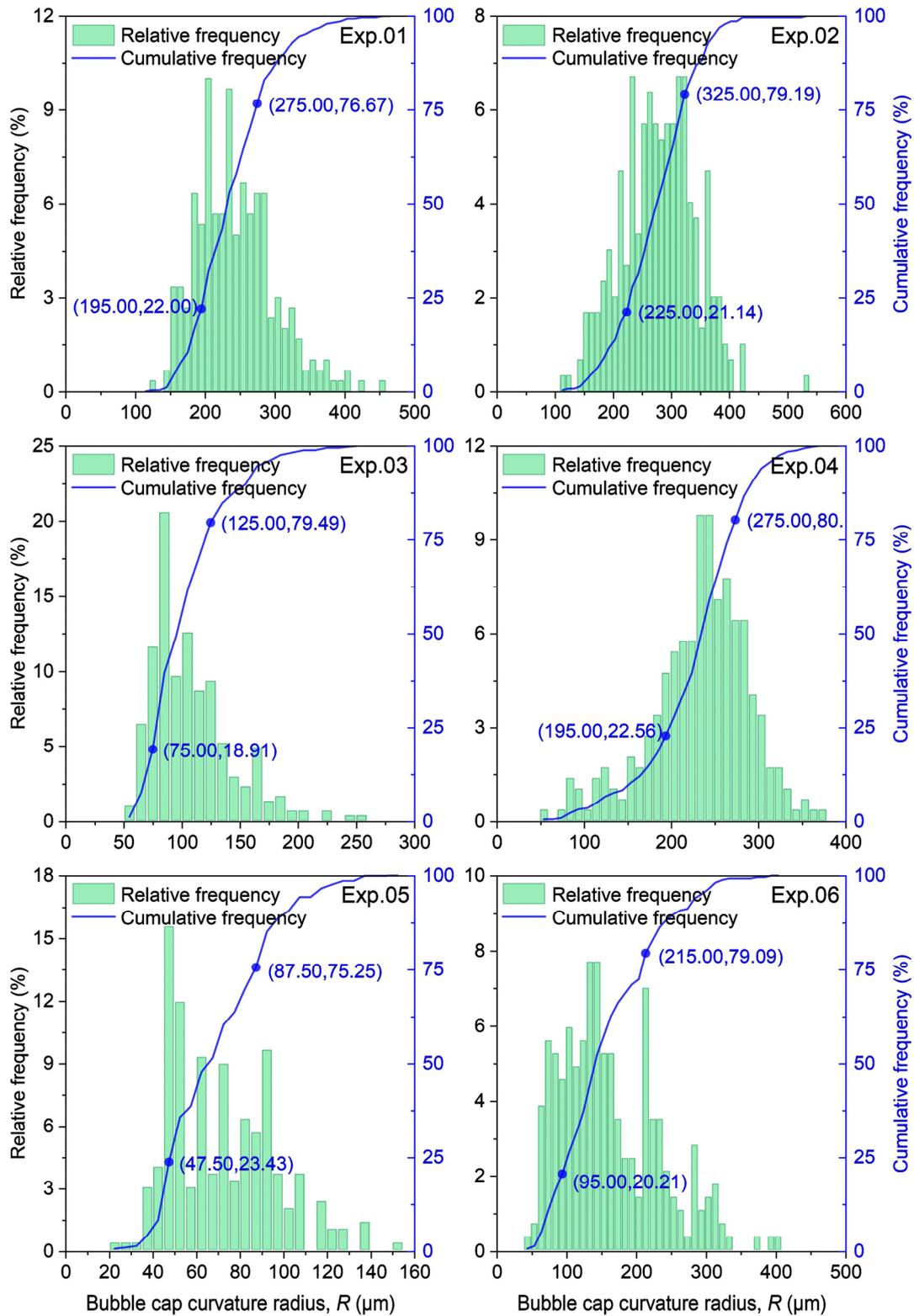

**Extended Data Fig. 9**

Particle size distributions of dried drops produced from the bubbles ($R \sim 900$ μm) with different residence time at the air flow rate of 30 mL/min.

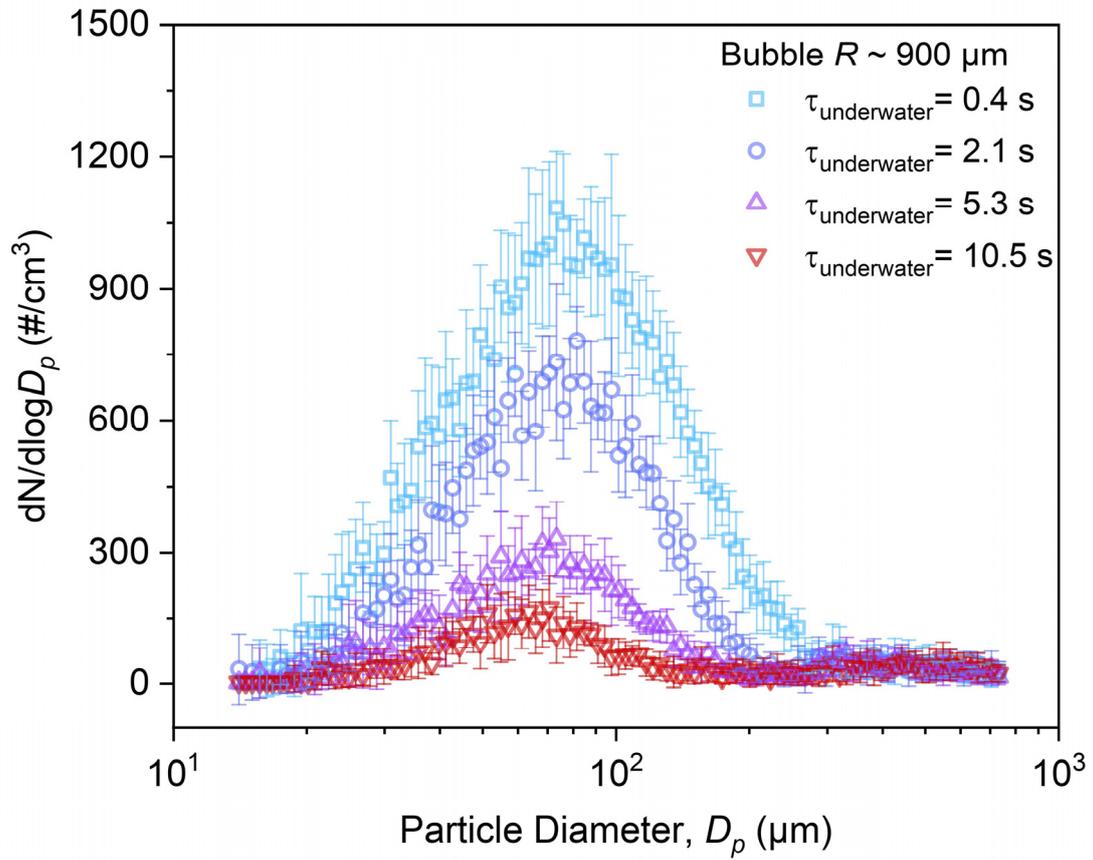

**Extended Data Fig. 10**

Schematic diagram of the experimental setup used to measure the air volume of bubbles rising to the water surface: A graduated cylinder filled with water is inverted and submerged in the water. As bubbles rise and enter the cylinder, they displace the water, causing the water level inside the cylinder to drop. This allows for an accurate measurement of the airflow rate of the bubble stream.

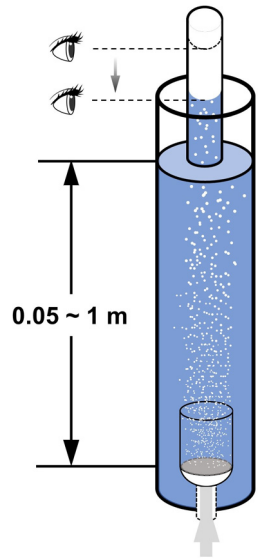

# SUPPLEMENTARY TABLES

**Supplementary Table 1.** Test plans for measurement of submicron drop concentrations from bubbles.

*A. for Fig. 1a*

| Serial | Frits of different porosities | Gas flow rate for bubble generation (sccm) | Supplementary gas flow rate (mL/min) | Aerosol Instrument |
|---|---|---|---|---|
| 1 | G2 | 2-20 | 3000 | 3775CPC |
| 2 | Needle_30G | 2-20 | 3000 | 3775CPC |
| 3 | Needle_24G | 6-50 | 3000 | 3775CPC |
| 4 | Needle_27G | 2-50 | 3000 | 3775CPC |
| 5 | Needle_25G | 5-50 | 3000 | 3775CPC |
| 6 | Needle_22G | 5-50 | 3000 | 3775CPC |

B. for Fig. 2b

| Bubble $R$ (μm) | Frits of different porosities | Supplementary gas flow rate (mL/min) | Airflow rate for bubble generation (sccm) | Height to the water surface (m) | Residence time $\tau_{underwater}$ (s) |
|---|---|---|---|---|---|
| ~ 900 | G2 | 3000 | 30 | 0.07 | 0.34 |
| | G2 | 3000 | 30 | 0.20 | 2.1 |
| | G2 | 3000 | 30 | 0.50 | 5.2 |
| | G2 | 3000 | 30 | 1.00 | 10.5 |
| ~ 950 | G2 | 3000 | 40 | 0.07 | 0.34 |
| | G2 | 3000 | 45 | 0.20 | 2.1 |
| | G2 | 3000 | 35 | 0.50 | 5.2 |
| | G2 | 3000 | 35 | 1.00 | 10.5 |
| ~ 1000 | G2 | 3000 | 50 | 0.07 | 0.34 |
| | G2 | 3000 | 50 | 0.20 | 2.1 |
| | G2 | 3000 | 50 | 0.50 | 5.2 |
| | G2 | 3000 | 50 | 1.00 | 10.5 |
| | G2 | 3000 | 25 | 2.00 | 21.0 |
| ~ 1050 | G2 | 3000 | 40 | 0.20 | 2.1 |
| | G2 | 3000 | 40 | 0.50 | 5.2 |
| | G2 | 3000 | 40 | 1.00 | 10.5 |
| | G2 | 3000 | 20 | 2.00 | 21.0 |

*C. for Fig. 2e*

| Bubble $R$ (μm) | Frits of different porosities | Airflow rate for bubble generation (sccm) | Supplementary gas flow rate (mL/min) | w/o bubble collision |
|---|---|---|---|---|
| ~ 1200 | Needle_30G | 5/5 | 1000 | No collision |
| | Needle_30G | 5/5 | 1000 | Collision |
| ~ 1500 | Needle_27G | 5/5 | 1000 | No collision |
| | Needle_27G | 5/5 | 1000 | Collision |
| ~ 2600 | Needle_22G | 10/10 | 1000 | No collision |
| | Needle_22G | 10/10 | 1000 | Collision |

D. for Fig. 4

| Bubble $R$ (μm) | Frits of different porosities | Supplementary gas flow rate (mL/min) | Airflow rate for bubble generation (sccm) | Height to the water surface (m) | Residence time $\tau_{underwater}$ (s) |
|---|---|---|---|---|---|
| ~ 1300 | Needle_32G | 3000 | 6 | 0.3 | 1.5 |
| ~ 910 | Needle_32G | 3000 | 6 | 3.0 | 10.0 |
| ~ 870 | Needle_34G | 3000 | 4 | 0.3 | 1.5 |
| ~ 840 | Needle_34G | 3000 | 4 | 3.0 | 10.0 |
| ~ 970 | G2 | 3000 | 25 | 0.07 | 0.4 |
| ~ 1000 | G2 | 1000 | 25 | 2.00 | 21.0 |
| ~ 290 | G3 | 3000 | 20 | 0.07 | 1.0 |
| ~ 280 | G3 | 3000 | 20 | 1.00 | 14.3 |
| ~ 130 | G4 | 3000 | 10 | 0.08 | 1.8 |
| ~ 190 | G4 | 3000 | 10 | 1.00 | 23.0 |
| ~ 80 | G5 | 3000 | 10 | 0.06 | 1.4 |
| ~ 190 | G5 | 3000 | 10 | 1.00 | 23.3 |
| ~ 2400 | Needle_24G | 1000 | 20/20 | 0.05 | 0.5 |
| ~ 2600 | Needle_22G | 1000 | 20/20 | 0.05 | 0.3 |
| ~ 3100 | Needle_22G | 1000 | 50/50 | 0.05 | 0.4 |
| ~ 3300 | Needle_22G 45° down | 1000 | 50/50 | 0.05 | 0.3 |

**Supplementary Table 2.** Test plan for measurement of bubble size ($R < \sim 400$ μm) for Fig. 4.

| Experiment number | Frits of different porosities | Airflow rate for bubble generation (sccm) | Height to the water surface (m) | Residence time $\tau_{underwater}$ (s) |
|---|---|---|---|---|
| Exp.01 | G3 | 20 | 0.07 | 1.0 |
| Exp.02 | G3 | 20 | 1.00 | 14.3 |
| Exp.03 | G4 | 10 | 0.08 | 1.8 |
| Exp.04 | G4 | 10 | 1.00 | 23.0 |
| Exp.05 | G5 | 10 | 0.06 | 1.4 |
| Exp.06 | G5 | 10 | 1.00 | 23.3 |

# SUPPLEMENTARY VIDEOS

**Supplementary Video 1.** High-speed visualization of bubble formation through a needle with an inner diameter of 0.21 mm under an airflow rate of 5 sccm. Captured at 20,000 fps and played back at 30 fps for clarity.

**Supplementary Video 2.** High-speed visualization of bubble formation through a needle with an inner diameter of 0.21 mm under an airflow rate of 10 sccm. Captured at 20,000 fps and played back at 30 fps for clarity.

**Supplementary Video 3.** Bubble formation from two separate needles (each with an inner diameter of 0.21 mm), oriented at a 45° downward angle, each with a flow rate of 5 sccm, representing the 'No collision' group. Captured at 20,000 fps and played back at 30 fps for clarity.

**Supplementary Video 4.** Bubble formation from two closely positioned needles (each with an inner diameter of 0.21 mm), oriented at a 45° downward angle, each with a flow rate of 5 sccm, representing the 'Collision' group. Captured at 20,000 fps and played back at 30 fps for clarity.

**Supplementary Video 5.** High-speed visualization of bubble formation through a needle with an inner diameter of 0.06 mm under an airflow rate of 5 sccm. Captured at 20,000 fps and played back at 30 fps for clarity.

**Supplementary Video 6.** Bursting of a film created by the collision of simulated air bubbles in test tube. Captured at 22,500 fps and played back at 25 fps for clarity.

**Supplementary Video 7.** Bursting of a film created by the collision of simulated $SF_6$

bubbles in test tube. Captured at 12,500 fps and played back at 25 fps for clarity.

**Supplementary Video 8.** Simulated bubble collision in a test tube (Fig. 3b). Captured at 240 fps and played back at 30 fps for clarity.

**Supplementary Video 9.** Underwater bubbles generated by a jet expelled from a 1/8-inch diameter tube at a flow rate of 368 mL/min. Captured at 20,000 fps and played back at 30 fps for clarity.